\documentclass[journal]{IEEEtran}

\usepackage{epsfig,amsmath,amssymb,epsf,amsthm,scalefnt,multirow,subfig}
\usepackage{xcolor}
\usepackage{float}
\usepackage{cite}
\usepackage{psfrag}

\newcommand{\argmax}{\mathop{\mathrm{argmax}}}
\newcommand{\argmin}{\mathop{\mathrm{argmin}}}

\newtheorem{lemma}{Lemma}
\newtheorem*{lemma*}{Lemma}

\newtheorem{corollary}{Corollary}

  \def\cC{{\mathcal{C}}}

 \def\cN{{\mathcal{N}}}  
  \def\cS{{\mathcal{S}}}

\def\argmin{\mathop{\mathrm{argmin}}}
\def\argmax{\mathop{\mathrm{argmax}}}

\def\b0{{\pmb{0}}} 

  \def\bc{{\mathbf{c}}} 
  \def\bg{{\mathbf{g}}} \def\bh{{\mathbf{h}}}
   
 \def\bn{{\mathbf{n}}}  
   \def\bt{{\mathbf{t}}}
\def\bu{{\mathbf{u}}}  \def\bw{{\mathbf{w}}} \def\bx{{\mathbf{x}}}
\def\by{{\mathbf{y}}}

   \def\bH{{\mathbf{H}}}
\def\bI{{\mathbf{I}}}

\begin{document}

\title{Distributed Reception with Spatial Multiplexing: MIMO Systems for the Internet of Things}

\author{Junil Choi, David J. Love, D. Richard Brown III, and Mireille Boutin\\
\thanks{Junil Choi, David J. Love, and Mireille Boutin are with the School of Electrical and Computer Engineering, Purdue University, West Lafayette, IN (e-mail: choi215@purdue.edu, djlove@purdue.edu, mboutin@purdue.edu).}
\thanks{D. Richard Brown III is with the Department of Electrical and Computer Engineering, Worcester Polytechnic Institute, Worcester, MA (e-mail: drb@wpi.edu).}
}


\maketitle

\begin{abstract}
The Internet of things (IoT) holds much commercial potential and could facilitate distributed multiple-input multiple-output (MIMO) communication in future systems.  We study a distributed reception scenario in which a transmitter equipped with multiple antennas sends multiple streams via spatial multiplexing to a large number of geographically separated single antenna receive nodes.  The receive nodes then quantize their received signals and forward the quantized received signals to a receive fusion center.  With global channel knowledge and forwarded quantized information from the receive nodes, the fusion center attempts to decode the transmitted symbols.  We assume the transmit vector consists of phase shift keying (PSK) constellation points, and each receive node quantizes its received signal with one bit for each of the real and imaginary parts of the signal to minimize the transmission overhead between the receive nodes and the fusion center.  Fusing this data is a non-trivial problem because the receive nodes cannot decode the transmitted symbols before quantization.  Instead, each receive node processes a single quantity, i.e., the received signal, regardless of the number of transmitted symbols.  We develop an optimal maximum likelihood (ML) receiver and a low-complexity zero-forcing (ZF)-type receiver at the fusion center.  Despite its suboptimality, the ZF-type receiver is simple to implement and shows comparable performance with the ML receiver in the low signal-to-noise ratio (SNR) regime but experiences an error rate floor at high SNR.  It is shown that this error floor can be overcome by increasing the number of receive nodes.  Hence, the ZF-type receiver would be a practical solution for distributed reception with spatial multiplexing in the era of the IoT where we can easily have a large number of receive nodes.
\end{abstract}


\section{Introduction}\label{sec1}

As more and more internet-enabled things are commonly used (e.g., computers, smartphones, tablets, home appliances, and more), the Internet of Things (IoT) will change the paradigm of communication systems \cite{IOT}.  In the IoT environment, devices could be used as distributed transmit and/or receive entities allowing massive distributed multiple-input multiple-output (MIMO) systems to be implemented.  Among many possible scenarios, we focus on distributed reception for wireless communication systems in this paper.

Distributed reception for wireless communication systems is used to provide reliable communication between a transmitter and a receive fusion center via the help of geographically separated receive nodes \cite{dist_detect2}.  Wireless channels between the transmitter and the multiple receive nodes are usually independent, resulting in increased diversity gain, and the fusion center estimates the transmitted data using processed information from the received nodes.  Distributed reception techniques are now adopted in 3GPP LTE-Advanced in the context of coordinated multipoint (CoMP) reception scenario for cellular systems \cite{comp1,comp2,comp3,comp4}.

There are strong similarities between distributed reception for wireless communication systems and wireless sensor networks (WSNs), where the former is aimed at data communications and the latter is more focused on environment classifications.  WSNs have been extensively studied in references such as \cite{wsn2,wsn3,wsn4,d_detect_overview1,d_detect_overview2,d_detect_overview3,dist_detect1,ribeiro1,ribeiro2,code_dist_detec1,code_dist_detec2,code_dist_detec3}. Many of the works on WSNs can be applied to distributed reception for wireless communication systems.  Instead of reusing ad-hoc processing and fusion rules as in \cite{code_dist_detec1,code_dist_detec2,code_dist_detec3} for WSNs when performing distributed reception of a communication signal, it was shown in \cite{coded_diversity1,coded_diversity2} that by adopting appropriate channel codes we can obtain simple, yet powerful processing and decoding rules that have good symbol error rate (SER) performance for a practical range of signal-to-noise ratios (SNRs).  Recently, \cite{brown} showed that even simple combining of hard decisions at the receive nodes can give performance within 2 dB of ideal receive beamforming for wireless communication systems.

However, most of the prior work on WSNs and distributed reception for wireless communication systems considered only detection/estimation of a single-dimensional parameter or single transmitted symbol.  To our knowledge, there are few papers that discuss multi-dimensional estimation problems.  A few exceptions can be found in \cite{d_detect_vec1,d_detect_vec2} which consider the estimation of a multi-dimensional vector in WSNs with additive noise at each sensor.

In this paper, we consider distributed MIMO communication systems where the transmitter is equipped with multiple antennas and simultaneously transmits independent data symbols chosen from a phase shift keying (PSK) constellation using spatial multiplexing to a set of geographically separated receive nodes deployed with a single receive antenna sent through independent fading channels.  Each receive node quantizes its received signal and forwards the quantized signal to the fusion center.  The fusion center then attempts to decode the transmitted data by exploiting the quantized signals from the receive nodes and global channel information.  This scenario is likely to become popular with the emergence of massive MIMO systems \cite{massive_mimo3} and IoT because base stations tend to be equipped with a large number of antennas in massive MIMO systems and we can easily have a large number of receive nodes in the IoT environment.

For practical purposes, we assume each receive node quantizes its received signal with one bit per real and imaginary part of the received signal to minimize the transmission overhead between the receive nodes and the fusion center.  Quantizer design is a non-trivial problem because the receive nodes are not able to decode the transmitted symbols due to the fact that each receive node has only one antenna \cite{sandell}.  Instead, each receive node quantizes a single quantity, i.e., the received signal, regardless of the number of transmitted symbols.  In this setup, we develop an optimal maximum likelihood (ML) receiver and a low-complexity zero-forcing (ZF)-type receiver assuming global channel knowledge at the fusion center.  The ML receiver outperforms the ZF-type receiver regardless of the number of receive nodes and SNR ranges.  However, the complexity of the ML receiver is excessive, especially when the number of transmitted symbols becomes large.  On the other hand, the ZF-type receiver can be easily implemented and gives comparable performance to that of the ML receiver when the SNR is low to moderate, although it suffers from an error rate floor when SNR is high.  The error rate floor of the ZF-type receiver can be easily mitigated by having more receive nodes.

When the SNR is high, the distributed reception problem is closely tied to work in quantized frame expansion.  Linear transformation and expansion by a frame matrix in the presence of coefficient quantization is thoroughly studied in \cite{frame_exp1,frame_exp2}.  A linear expansion method, which is similar to our ZF-type receiver, and its performance in terms of the mean-squared error (MSE) are analyzed based on the properties of a frame matrix.  An advanced non-linear expansion method relying on linear programming is also studied.  The major difference compared to our problem setting is that \cite{frame_exp1,frame_exp2} do not assume any additive noise before quantization, while our scenario considers a fading channel (which corresponds to a frame matrix in frame expansion) with additive noise prior to quantization at the receive nodes.  We rely on some of the analytical results from \cite{frame_exp1} for evaluating and modifying the ZF-type receiver later.


The rest of the paper is organized as follows.  In Section \ref{sec2}, we define our system model.  The ML and ZF-type receivers are proposed and their characteristics are compared in Section \ref{receiver}.  Simulation results that evaluate the proposed receivers are shown in Section~\ref{simulation}, and conclusions follow in Section~\ref{conclusion}.

\noindent \textit{\textbf{Notation}}: Lower and upper boldface symbols denote column vectors and matrices, respectively.  $\|\bx\|$ represents the two-norm of a vector $\bx$, and $(\cdot)^T$, $(\cdot)^H$, $(\cdot)^\dagger$ are used to denote transpose, Hermitian transpose, and pseudo inverse of their argument, respectively.  $\mathrm{Re}(c)$ and $\mathrm{Im}(c)$ denote the real and complex part of a complex number $c$, respectively.  $\mathbf{0}_m$ represents an $m\times 1$ all zero vector, $\mathbf{1}_m$ denotes an $m\times 1$ all one vector, and $\bI_m$ is used for $m \times m$ identity matrix.  $\mathcal{I}(\cdot)$ is used as the indicator function which equals to one if the argument is true and zero otherwise, and ${\rm Pr}(A)$ denotes the probability of event $A$.

\section{System Model}\label{sec2}
We consider a network consisting of a transmitter with $N_t$ antennas, communicating with a receive fusion center that is connected to $K$ geographically separated, single antenna receive nodes.  The transmitter tries to send $N_t$ independent data symbols simultaneously by spatial multiplexing\footnote{The transmitter also can send a number of symbols smaller than $N_t$ by adopting precoding or antenna selection, which is out of scope of this paper.} to the fusion center via the help of the receive nodes.  The received signal at the $k$-th receive node is given as
\begin{equation}\label{input_output}
  y_k=\sqrt{\frac{\rho}{N_t}}\bh_k^H\bx+n_k,\quad k=1,\cdots,K
\end{equation}
where $\rho$ is the transmit SNR, $\bh_k\in \mathbb{C}^{N_t}$ is the independent and identically distributed (i.i.d.) Rayleigh fading channel vector between the transmitter and the $k$-th receive node, $n_k$ is complex additive white Gaussian noise (AWGN) distributed as $\cC\cN(0,1)$ at the $k$-th node, and $\bx=[x_1,\cdots,x_{N_t}]^T$ is the transmitted signal vector where $x_i\in \cS$ is from a standard $M$-ary PSK constellation
\begin{equation*}
  \cS = \{s_1,\cdots,s_M\}\subset \mathbb{C},
\end{equation*}
which satisfies $|s_m|^2=1$ for all $m$ and $\|\bx\|^2=N_t$.  We assume that $x_i$ is drawn from $\cS$ with equal probabilities.  The input-output relation in \eqref{input_output} can be also written as
\begin{equation*}
  \by = \sqrt{\frac{\rho}{N_t}}\bH\bx+\bn
\end{equation*}
where
\begin{align*}
\nonumber  \by&=\begin{bmatrix}y_1 & {y}_2 & \cdots & {y}_K\end{bmatrix}^T,\\
\nonumber  \bn&=\begin{bmatrix}n_1 & {n}_2 & \cdots & {n}_K\end{bmatrix}^T,\\
  \bH&=\begin{bmatrix}\bh_1 & \bh_2 & \cdots & \bh_K\end{bmatrix}^H.
\end{align*}
We further assume that the fusion center can access the full knowledge of $\bh_k$ for all $k$.  Extending our framework to partial or no channel knowledge at the fusion center or other constellations such as quadrature amplitude modulation (QAM) is clearly possible and is an interesting future research topics.

If the fusion center has full knowledge of $y_k$ for all $k$, then the optimal receiver is given as
\begin{equation*}
  \hat{\bx}_{\mathrm{opt}}=\argmin_{\bx'\in \cS^{N_t}}\left\lVert \by-\sqrt{\frac{\rho}{N_t}}\bH\bx'\right\rVert^2
\end{equation*}
where $\cS^{n}$ is the cartesian product of $\cS$ of order $n$.  However, we are interested in the scenario when each receive node \textit{quantizes} its received signal and conveys the quantized received signal, $\hat{y}_k$, to the fusion center.  Therefore, the fusion center needs to have other approaches to decode the transmitted symbols in our problem.

We assume $\hat{y}_k$ can be forwarded from the $k$-th receive node to the fusion center without any error.  This assumption would be reasonable because the receive nodes and the fusion center are usually connected by a very high-rate link or located near each other in practice.  We further assume that the forward link transmission and the LAN are operated on different time or frequency resources to prevent interference between the two.

To make the problem practical, we assume that the receive nodes only can perform very simple operation, i.e., they do not decode the transmitted vector $\bx$ but instead simply quantize $y_k$ directly.  Moreover, to minimize the data transmission overhead from the receive nodes to the fusion center, we assume each receive node quantizes $y_k$ using two bits, i.e., one bit for each of the real and imaginary parts of $y_k$.  Thus, the quantized received signal $\hat{y}_k$ can be written as
\begin{equation*}
\hat{y}_k = \mathrm{sgn}(\mathrm{Re}(y_k)-\tau_{\mathrm{Re},k})+j\left(\mathrm{sgn}(\mathrm{Im}(y_k)-\tau_{\mathrm{Im},k})\right)
\end{equation*}
where $\mathrm{sgn}(\cdot)$ is the sign function defined as
\begin{equation*}
\mathrm{sgn}(x) = \begin{cases}1& \text{if } x\geq 0 \\
-1 & \text{if } x<0\end{cases},
\end{equation*}
and $\tau_{\mathrm{Re},k}$ and $\tau_{\mathrm{Im},k}$ are quantization thresholds of the real and imaginary parts of $y_k$ at user $k$, respectively.

With a given realization of $\bh_k$, we consider the simple, yet effective, thresholds
\begin{align*}
  \tau_{\mathrm{Re},k} &= \mathrm{E}\left[\mathrm{Re}(y_k)\right] = \mathrm{E}\left[\sqrt{\frac{\rho}{N_t}}\mathrm{Re}(\bh_k^H\bx)+\mathrm{Re}(n_k)\right]=0,\\
  \tau_{\mathrm{Im},k} &= \mathrm{E}\left[\mathrm{Im}(y_k)\right] = \mathrm{E}\left[\sqrt{\frac{\rho}{N_t}}\mathrm{Im}(\bh_k^H\bx)+\mathrm{Im}(n_k)\right]=0,
\end{align*}
where equalities are based on the assumption that $n_k$ is distributed as $\cC\cN(0,1)$, or equivalently $\mathrm{Re}(n_k)$ and $\mathrm{Im}(n_k)$ are independent and both distributed as $\cN(0,\frac{1}{2})$, and the entries of $\bx$ are independently drawn from $\cS^{N_t}$ with equal probabilities, which gives $\mathrm{E}[\mathrm{Re}(\bc^T\bx)]=0$ and $\mathrm{E}[\mathrm{Im}(\bc^T\bx)]=0$ for an arbitrary combining vector $\bc\in\mathbb{C}^{N_t}$.  Although simple, these thresholds are consistent with the optimal threshold design studied in \cite{d_detect_vec1} in an average sense.  We assume the quantization thresholds $\tau_{{\rm Re},k}=0$ and $\tau_{{\rm Im},k}=0$ for the remainder of this paper.

\begin{figure}[t]
  \centering
  \psfrag{h}{$\bH=\begin{bmatrix}\bh_1 & \bh_2 & \cdots \bh_K\end{bmatrix}^H$}
  \psfrag{x}{$\hat{y}_1$}
  \psfrag{y}{$\hat{y}_2$}
  \psfrag{z}{$\hat{y}_K$}
  \includegraphics[width=0.8\columnwidth]{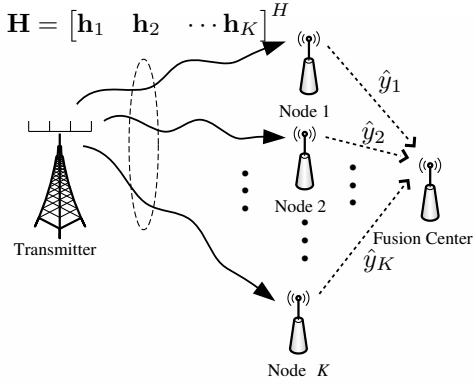}\\
  \caption{The conceptual figure of distributed reception with multiple antennas at the transmitter.  Each receive node is equipped with a single receive antenna.}\label{concept}
\end{figure}

Once the fusion center receives $\hat{y}_k$ from all receive nodes, it attempts to decoded the transmitted data symbols $\bx$ using the forwarded information and channel knowledge.  We define
\begin{align*}
  \hat{\by}=\begin{bmatrix}\hat{y}_1 & \hat{y}_2 & \cdots & \hat{y}_K\end{bmatrix}^T
\end{align*}
which is useful in Section \ref{zf_rec}.  The conceptual explanation of the scenario is depicted in Fig. \ref{concept}.

\section{Decoding Rules at Fusion Center}\label{receiver}
With the knowledge of $\bH$ and $\hat{\by}$ at the fusion center, we can implement different kinds of receivers considering complexity and performance.  We first develop an optimal ML receiver and low-complexity ZF-type receiver.  Then, we discuss the performance of receivers regarding system parameters such as $\rho$ and $K$.

\subsection{ML receiver}\label{ml_receiver_sec}
We convert the problem of interest to the real domain to facilitate analysis.  This can be done by defining $\bH_{\mathrm{R},k}\in \mathbb{R}^{2 \times 2 N_t}$, $\bx_{\mathrm{R}}\in \mathbb{R}^{2 N_t}$ and $\bn_{\mathrm{R},k}\in \mathbb{R}^{2}$ as
\begin{align*}
\bH_{\mathrm{R},k}&=\begin{bmatrix} \mathrm{Re}(\bh_k^T) & \mathrm{Im}(\bh_k^T) \\ -\mathrm{Im}(\bh_k^T) & \mathrm{Re}(\bh_k^T)\end{bmatrix}=\begin{bmatrix}\bh_{\mathrm{R},k,1}^T \\ \bh_{\mathrm{R},k,2}^T \end{bmatrix},\\
\bx_{\mathrm{R}}&=\begin{bmatrix} \mathrm{Re}(\bx) \\ \mathrm{Im}(\bx)\end{bmatrix},\\
\bn_{\mathrm{R},k}&=\begin{bmatrix} \mathrm{Re}(n_k) \\ \mathrm{Im}(n_k)\end{bmatrix}
\end{align*}
where
\begin{equation*}
  \bh_{\mathrm{R},k,1}=\begin{bmatrix}\mathrm{Re}(\bh_k) \\ \mathrm{Im}(\bh_k)	\end{bmatrix},\quad \bh_{\mathrm{R},k,2}=\begin{bmatrix}-\mathrm{Im}(\bh_k) \\ \mathrm{Re}(\bh_k)	\end{bmatrix}.
\end{equation*}
Then, the received signal $y_k$ also can be rewritten in the real domain as
\begin{align*}
\by_{\mathrm{R},k}=\begin{bmatrix} y_{\mathrm{R},k,1} \\ y_{\mathrm{R},k,2}\end{bmatrix}=\begin{bmatrix} \mathrm{Re}(y_k) \\ \mathrm{Im}(y_k)\end{bmatrix}=\sqrt{\frac{\rho}{N_t}}\bH_{\mathrm{R},k} \bx_{\mathrm{R}} + \bn_{\mathrm{R},k},
\end{align*}
and the vectorized version of the quantized $\hat{y}_k$ in the real domain is given as
\begin{equation}
\hat{\by}_{\mathrm{R},k}=\begin{bmatrix} \hat{y}_{\mathrm{R},k,1} \\ \hat{y}_{\mathrm{R},k,2}\end{bmatrix}=\begin{bmatrix} \mathrm{sgn}(\mathrm{Re}(y_k)) \\ \mathrm{sgn}(\mathrm{Im}(y_k))\end{bmatrix}.\label{real_quantized}
\end{equation}

Once the fusion center receives $\hat{\by}_{\mathrm{R},k}$ from all receive nodes, it generates the \textit{sign-refined} channel matrix $\widetilde{\bH}_{\mathrm{R},k}$ according to
\begin{equation*}
\widetilde{\bH}_{\mathrm{R},k}=\begin{bmatrix}\widetilde{\bh}_{\mathrm{R},k,1}^T \\ \widetilde{\bh}_{\mathrm{R},k,2}^T \end{bmatrix}
\end{equation*}
where $\widetilde{\bh}_{\mathrm{R},k,i}$ is defined as
\begin{align}\label{sign_refine}
  \widetilde{\bh}_{\mathrm{R},k,i}=\hat{y}_{\mathrm{R},k,i}\bh_{\mathrm{R},k,i}.
\end{align}
Because $\hat{y}_{\mathrm{R},k,i}$ is $\pm 1$, \eqref{sign_refine} can be considered as a sign refinement of $\bh_{\mathrm{R},k,i}$.  We let $\cS_{\mathrm{R}}$ be
\begin{equation*}
  \cS_{\mathrm{R}}=\left\{\begin{bmatrix}\mathrm{Re}(s_1) \\ \mathrm{Im}(s_1)\end{bmatrix},\cdots,\begin{bmatrix}\mathrm{Re}(s_M) \\ \mathrm{Im}(s_M)\end{bmatrix}\right\}
\end{equation*}
where $M$ is the size of the constellation $\cS$.  We also define two sets $\mathcal{P}_{\bx_{\mathrm{R}}}$ and $\mathcal{N}_{\bx_{\mathrm{R}}}$ as
\begin{align*}
  \mathcal{P}_{\bx_{\mathrm{R}}} & =\left\{(k,i):~y_{\mathrm{R},k,i}\geq 0~ \text{given}~ \bx_{\mathrm{R}}~\text{is sent.}\right\}, \\
  \mathcal{N}_{\bx_{\mathrm{R}}} & =\left\{(k,i):~y_{\mathrm{R},k,i}< 0~ \text{given}~ \bx_{\mathrm{R}}~\text{is sent.} \right\}
\end{align*}
for $k\in\left\{1,\cdots,K\right\}$ and $i\in\left\{1,2\right\}$.

With these definitions, we can define a likelihood function as
\begin{align}
\nonumber L(\bx_{\mathrm{R}}')&=\mathrm{Pr}\left(\left.\sqrt{\frac{\rho}{N_t}}\bh_{\mathrm{R},k,i}^T\bx_{\mathrm{R}}'+n_{\mathrm{R},k,i}\geq 0\right|\forall (k,i)\in\mathcal{P}_{\bx_{\mathrm{R}}'}\right)\\
\nonumber &  \quad \cdot \mathrm{Pr}\left(\left.\sqrt{\frac{\rho}{N_t}}\bh_{\mathrm{R},k,i}^T\bx_{\mathrm{R}}'+n_{\mathrm{R},k,i}< 0\right|\forall(k,i)\in\mathcal{N}_{\bx_{\mathrm{R}}'}\right)\\
\nonumber &=\mathrm{Pr}\left(\left.\sqrt{\frac{\rho}{N_t}}\bh_{\mathrm{R},k,i}^T\bx_{\mathrm{R}}'\geq -n_{\mathrm{R},k,i}\right|\forall(k,i)\in\mathcal{P}_{\bx_{\mathrm{R}}'}\right)\\
\nonumber &  \quad \cdot \mathrm{Pr}\left(\left.\sqrt{\frac{\rho}{N_t}}\bh_{\mathrm{R},k,i}^T\bx_{\mathrm{R}}'<-n_{\mathrm{R},k,i}\right|\forall(k,i)\in\mathcal{N}_{\bx_{\mathrm{R}}'}\right)\\
\nonumber &\stackrel{(a)}{=}\mathrm{Pr}\left(\left.\sqrt{\frac{\rho}{N_t}}\widetilde{\bh}_{\mathrm{R},k,i}^T\bx_{\mathrm{R}}'\geq -n_{\mathrm{R},k,i}\right|\forall(k,i)\in\mathcal{P}_{\bx_{\mathrm{R}}'}\right)\\
\nonumber &  \quad \cdot \mathrm{Pr}\left(\left.\sqrt{\frac{\rho}{N_t}}\widetilde{\bh}_{\mathrm{R},k,i}^T\bx_{\mathrm{R}}'\geq n_{\mathrm{R},k,i}\right|\forall(k,i)\in\mathcal{N}_{\bx_{\mathrm{R}}'}\right)\\
\nonumber &\stackrel{(b)}{=}\mathrm{Pr}\left(\left.\sqrt{\frac{\rho}{N_t}}\widetilde{\bh}_{\mathrm{R},k,i}^T\bx_{\mathrm{R}}'\geq -n_{\mathrm{R},k,i}\right|\forall(k,i)\in\mathcal{P}_{\bx_{\mathrm{R}}'}\right)\\
\nonumber &  \quad \cdot \mathrm{Pr}\left(\left.\sqrt{\frac{\rho}{N_t}}\widetilde{\bh}_{\mathrm{R},k,i}^T\bx_{\mathrm{R}}'\geq -n_{\mathrm{R},k,i}\right|\forall(k,i)\in\mathcal{N}_{\bx_{\mathrm{R}}'}\right)\\
\nonumber &\stackrel{(c)}{=}\prod_{i=1}^{2}\prod_{k=1}^{K}\Phi\left(\sqrt{\frac{2\rho}{N_t}}\widetilde{\bh}_{\mathrm{R},k,i}^T\bx_{\mathrm{R}}'\right)
\end{align}
where $\Phi(t)=\int_{-\infty}^{t}\frac{1}{\sqrt{2\pi}}e^{-\frac{\tau^2}{2}}d\tau$, $(a)$ is based on the sign refinement in \eqref{sign_refine}, $(b)$ is because $n_{\mathrm{R},k,i}$ and $-n_{\mathrm{R},k,i}$ have the same distribution (or the same probability density function) such that $\mathrm{Pr}\left(c \geq n_{\mathrm{R},k,i}\right)=\mathrm{Pr}\left(c \geq -n_{\mathrm{R},k,i}\right)$ for an arbitrary constant $c$ and $(c)$ comes from the fact that $n_{\mathrm{R},k,i}$ is independent for all $k$ and $i$ and from distribution $\cN\left(0,\frac{1}{2}\right)$.  Then, the ML receiver is given as\footnote{A similar ML receiver is also derived in \cite{d_detect_vec1}.}
\begin{align}\label{obj_func}
\hat{\bx}_{\mathrm{R,ML}}=\argmax_{\bx_{\mathrm{R}}'\in\cS_{\mathrm{R}}^{N_t}} \prod_{i=1}^{2}\prod_{k=1}^{K}\Phi\left(\sqrt{\frac{2\rho}{N_t}}\widetilde{\bh}_{\mathrm{R},k,i}^T\bx_{\mathrm{R}}'\right).
\end{align}

The complexity of the exhaustive search of the ML receiver increases exponentially with the number of transmit symbols in spatial multiplexing, i.e., we need to search over $M^{N_t}$ elements.  Therefore, in practice, it is desired to implement a low complexity receiver for large numbers of transmit antennas.


\noindent \textbf{Remark 1:} If the number of receive nodes $K$ is less than the number of transmit antennas $N_t$, then the decoding performance at the fusion center would be very poor.  This situation will likely not hold for our problem setting because we can easily have $K \gg N_t$ based on the IoT environment.

\noindent \textbf{Remark 2:} Instead of quantizing both the real and imaginary parts of the received signal at each node, we can have the same performance on average by quantizing and forwarding only the real \textit{or} imaginary part of the received signal with twice the number of received nodes.  This is based on the assumption that the real and imaginary parts of the noise $n_k$ are i.i.d. for all $k$.

\subsection{Low-complexity zero-forcing-type receiver}\label{zf_rec}
Before proposing our ZF-type receiver, we first state the following lemma which establishes the theoretical foundation of our receiver.

\begin{lemma}\label{likeli_upper}
Define a matrix $\widetilde{\bH}_{\mathrm{R,S}}\in \mathbb{R}^{2K\times 2N_t}$ by stacking $\widetilde{\bH}_{\mathrm{R},k}$ as
\begin{align}\label{H_tilde_def}
  \widetilde{\bH}_{\mathrm{R,S}}&=\begin{bmatrix}\widetilde{\bH}_{\mathrm{R},1}^T & \widetilde{\bH}_{\mathrm{R},2}^T \cdots & \widetilde{\bH}_{\mathrm{R},K}^T\end{bmatrix}^T,
\end{align}
and let $\bt(\bx_{\mathrm{R}}')$ be
\begin{align*}
  \bt(\bx_{\mathrm{R}}') &= \begin{bmatrix}t_1(\bx_{\mathrm{R}}') & t_2(\bx_{\mathrm{R}}') & \cdots & t_{2K}(\bx_{\mathrm{R}}')\end{bmatrix}^T,\\
t_{\ell}(\bx_{\mathrm{R}}') &= \widetilde{\bh}_{\mathrm{R},k,i}^T\bx_{\mathrm{R}}'
\end{align*}
where $\ell=2(k-1)+i$ for $k=1,\cdots,K$ and $i=1,2$.  Note that $\|\bx_{\mathrm{R}}'\|^2=N_t$ based on the PSK constellation assumption.  Then the likelihood function $L(\bx_{\mathrm{R}}')$ is upper bounded as
\begin{align*}
  L(\bx_{\mathrm{R}}')&= \prod_{i=1}^{2}\prod_{k=1}^{K}\Phi\left(\sqrt{\frac{2\rho}{N_t}}\widetilde{\bh}_{\mathrm{R},k,i}^T\bx_{\mathrm{R}}'\right)\\
  &=\prod_{\ell=1}^{2K}\Phi\left(\sqrt{\frac{2\rho}{N_t}}t_\ell(\bx_{\mathrm{R}}')\right)\\
  &\leq \prod_{\ell=1}^{2K}\Phi\left(\sqrt{\frac{\rho}{K}}\|\widetilde{\bH}_{\mathrm{R,S}}\|_{\mathrm{A}}\right)
\end{align*}
when $t_{\ell}(\bx_{\mathrm{R}}')= \sqrt{\frac{N_t}{2K}}\|\widetilde{\bH}_{\mathrm{R,S}}\|_{\mathrm{A}}$ for all $\ell$ where $\|\cdot\|_{\mathrm{A}}$ is an arbitrary matrix norm that is consistent with the vector two-norm.
\end{lemma}
\begin{IEEEproof}
To prove Lemma \ref{likeli_upper}, we derive an upper bound of the maximum of $L(\bx_{\mathrm{R}}')$ with the relaxed constraint $\bx_{\mathrm{R}}'\in \mathbb{R}^{2N_t}$ instead of $\bx_{\mathrm{R}}'\in \cS_{\mathrm{R}}^{N_t}$.  Note that the norm constraint $\|\bx_{\mathrm{R}}'\|^2=N_t$ still holds.  With the definitions of $t_{\ell}(\bx_{\mathrm{R}}')$ and $\bt(\bx_{\mathrm{R}}')$, we have
\begin{align}
&\nonumber  \max_{\substack{\bx_{\mathrm{R}}'\in \mathbb{R}^{2N_t},\\ \|\bx_{\mathrm{R}}'\|^2=N_t}} L(\bx_{\mathrm{R}}')\\
&\quad = \max_{\substack{\bx_{\mathrm{R}}'\in \mathbb{R}^{2N_t},\\ \|\bx_{\mathrm{R}}'\|^2=N_t}} \prod_{i=1}^{2}\prod_{k=1}^{K}\Phi\left(\sqrt{\frac{2\rho}{N_t}}\widetilde{\bh}_{\mathrm{R},k,i}^T\bx_{\mathrm{R}}'\right)\\
&\quad \leq \max_{\substack{\bt(\bx_{\mathrm{R}}')\in \mathbb{R}^{2K},\\ \|\bt(\bx_{\mathrm{R}}')\|^2\leq N_t\|\widetilde{\bH}_{\mathrm{R,S}}\|_{\mathrm{A}}^2,\\ t_{\ell}(\bx_{\mathrm{R}}')>0, \forall k }} \prod_{\ell=1}^{2K}\Phi\left(\sqrt{\frac{2\rho}{N_t}}t_{\ell}(\bx_{\mathrm{R}}') \right)\label{likeli_upper_proof1}\\
  &\quad =\max_{\substack{\bt(\bx_{\mathrm{R}}')\in \mathbb{R}^{2K},\\ \|\bt(\bx_{\mathrm{R}}')\|^2= N_t\|\widetilde{\bH}_{\mathrm{R,S}}\|_{\mathrm{A}}^2,\\ t_{\ell}(\bx_{\mathrm{R}}')>0, \forall k }} \prod_{\ell=1}^{2K}\Phi\left(\sqrt{\frac{2\rho}{N_t}}t_{\ell}(\bx_{\mathrm{R}}') \right)\label{likeli_upper_proof}
\end{align}
where \eqref{likeli_upper_proof1} is based on the facts that
\begin{align*}
\|\widetilde{\bH}_{\mathrm{R,S}}\bx_{\mathrm{R}}'\|^2\leq \|\widetilde{\bH}_{\mathrm{R,S}}\|_{\mathrm{A}}^2 \|\bx_{\mathrm{R}}'\|^2 =N_t \|\widetilde{\bH}_{\mathrm{R,S}}\|_{\mathrm{A}}^2
\end{align*}
and $\Phi(a)\geq \Phi(b)$ for $a\geq b$.  Note that the inequality constraint on $\|\bt(\bx_{\mathrm{R}}')\|^2$ in \eqref{likeli_upper_proof1} is changed to the equality constraint in \eqref{likeli_upper_proof}.

The objective function in \eqref{likeli_upper_proof} is trivially bounded by one; however, there is a certain maximum point in our problem because of the norm constraint of $\bt(\bx_{\mathrm{R}}')=N_t\|\widetilde{\bH}_{\mathrm{R,S}}\|_{\mathrm{A}}^2$.  Let $g_{\ell}=\sqrt{\frac{2\rho}{N_t}}t_{\ell}(\bx_{\mathrm{R}}')$ and $\bg=\begin{bmatrix}g_1 & g_2 & \cdots & g_{2K}\end{bmatrix}^T$.  Instead of finding the solution for \eqref{likeli_upper_proof} directly, we first find a local extrema of
\begin{align}
\nonumber  \log \left[\prod_{\ell=1}^{2K}\Phi\left(\sqrt{\frac{2\rho}{N_t}}t_{\ell}(\bx_{\mathrm{R}}') \right)\right]&=\sum_{\ell=1}^{2K}\log \left[\Phi\left(\sqrt{\frac{2\rho}{N_t}}t_{\ell}(\bx_{\mathrm{R}}') \right)\right]\\
  &=\sum_{\ell=1}^{2K}\log\Phi\left(g_{\ell} \right)\label{tangential_object}
\end{align}
by looking at the point at which the tangential derivatives to the circle $\|\bg\|^2= 2\rho \|\widetilde{\bH}_{\mathrm{R,S}}\|_{\mathrm{A}}^2$ are equal to zero.\footnote{Because our searching space is restricted to the circle $\|\bg\|^2= 2\rho \|\widetilde{\bH}_{\mathrm{R,S}}\|_{\mathrm{A}}^2$, the point where the tangential derivatives equal to zero is a local extrema of the objective function.}  The tangential derivatives of \eqref{tangential_object} are given by
\begin{align*}
  &\left(g_{n}\frac{\partial}{\partial g_m}-g_{m}\frac{\partial}{\partial g_{n}}\right)\sum_{\ell=1}^{2K}\log \Phi\left(g_{\ell} \right)\\
  &\qquad\qquad\qquad =g_{n}\frac{\Phi'\left(g_m \right)}{\Phi\left(g_m \right)}-g_{m}\frac{\Phi'\left(g_{n} \right)}{\Phi\left(g_{n} \right)}
\end{align*}
for $n,m=1,2,\cdots,2K$ and $n\neq m$.  Setting the tangential derivatives equal to zero, we obtain the equations
\begin{align*}
g_{n}\frac{\Phi'\left(g_m \right)}{\Phi\left(g_m \right)}=g_{m}\frac{\Phi'\left(g_{n} \right)}{\Phi\left(g_{n} \right)}
\end{align*}
or equivalently,
\begin{align}
\frac{1}{g_{m}}\frac{\Phi'\left(g_m \right)}{\Phi\left(g_m \right)}=\frac{1}{g_{n}}\frac{\Phi'\left(g_{n} \right)}{\Phi\left(g_{n} \right)}\label{tang_eq}
\end{align}
for $n,m=1,2,\cdots,2K$ and $n\neq m$ because $g_{\ell}>0$ for all $\ell$.  Clearly, this system of equations is satisfied when $g_{n}=g_m$ for all $n,m=1,2,\cdots,2K$.  Under the constraint $\|\bg\|^2= 2\rho \|\widetilde{\bH}_{\mathrm{R,S}}\|_{\mathrm{A}}^2$, one possible solution point is given as
\begin{equation}\label{g_soultion}
  g_{\ell}=\sqrt{\frac{\rho }{K}}\|\widetilde{\bH}_{\mathrm{R,S}}\|_{\mathrm{A}}
\end{equation}
for all $\ell$.  Note that the point in \eqref{g_soultion} is the only solution for \eqref{tang_eq} because
\begin{equation*}
  G(s)=\frac{1}{s}\Phi'(s)\frac{1}{\Phi(s)}
\end{equation*}
is a product of three functions that are strictly monotonically decreasing with $s\in(0,\infty)$, and thus $G(s)$ is also strictly monotonically decreasing with $s$.

Because $t_{\ell}(\bx_{\mathrm{R}}')=\sqrt{\frac{N_t}{2\rho}}g_{\ell}$, the point
\begin{equation*}
  t_{\ell}(\bx_{\mathrm{R}}') = \sqrt{\frac{N_t}{2K}}\|\widetilde{\bH}_{\mathrm{R,S}}\|_{\mathrm{A}}
\end{equation*}
for $\ell=1,\cdots,2K$ is the only extreme point of the objective function in \eqref{likeli_upper_proof}.  We can show that the extreme point is indeed the maximum point of \eqref{likeli_upper_proof} by using the lemma in Appendix \ref{phi_lemma}.
\end{IEEEproof}

Lemma \ref{likeli_upper} states that when $\bt(\bx_{\mathrm{R}}')=\sqrt{\frac{N_t}{2K}}\|\widetilde{\bH}_{\mathrm{R,S}}\|_{\mathrm{A}}\mathbf{1}_{2K}$, it maximizes the likelihood function with the norm constraint $\|\bt(\bx_{\mathrm{R}}')\|^2= N_t\|\widetilde{\bH}_{\mathrm{R,S}}\|_{\mathrm{A}}^2$.  From the fact that
\begin{equation*}
  \bt(\bx_{\mathrm{R}}') = \widetilde{\bH}_{\mathrm{R,S}}\bx'_{\mathrm{R}},
\end{equation*}
the vector $\check{\bx}_{\mathrm{R}}$, which is given as
\begin{equation*} \check{\bx}_{\mathrm{R}}=\widetilde{\bH}_{\mathrm{R,S}}^{\dagger}\bt(\bx_{\mathrm{R}}')=\sqrt{\frac{N_t}{2K}}\|\widetilde{\bH}_{\mathrm{R,S}}\|_{\mathrm{A}}\widetilde{\bH}_{\mathrm{R,S}}^{\dagger}\mathbf{1}_{2K},
\end{equation*}
would be a reasonable estimate for the transmitted vector.  Note that the norm of $\check{\bx}_{\mathrm{R}}$ may not be $N_t$ anymore; however, the normalization term does not have any impact on PSK symbol decisions.

To implement this receiver in terms of the quantized received signals, let $\hat{\by}_{\mathrm{R}}$ be
\begin{equation*}
  \hat{\by}_{\mathrm{R}}=\begin{bmatrix}\hat{\by}_{\mathrm{R},1}^T & \hat{\by}_{\mathrm{R},2}^T & \cdots & \hat{\by}_{\mathrm{R},K}^T\end{bmatrix}^T
\end{equation*}
where $\hat{\by}_{\mathrm{R},k}$ is defined in \eqref{real_quantized}.  It is easy to show by using the relation between $\bH_{\mathrm{R,S}}$ and $\widetilde{\bH}_{\mathrm{R,S}}$ (or between their rows given in \eqref{sign_refine}) that
\begin{equation*}
  \widetilde{\bH}^{\dagger}_{\mathrm{R,S}}\mathbf{1}_{2K} = \bH^{\dagger}_{\mathrm{R,S}}\hat{\by}_{\mathrm{R}}
\end{equation*}
because the $i$-th row of $\widetilde{\bH}_{\mathrm{R,S}}$ is the same as that of $\bH_{\mathrm{R,S}}$ with the sign adjustment by the sign of the $i$-th element of $\hat{\by}_{\mathrm{R}}$.  Based on these observations, we propose a ZF-type receiver at the fusion center, i.e., the fusion center generates $\check{\bx}_{\mathrm{R,ZF}}\in \mathbb{R}^{2N_t}$ as
\begin{equation}\label{zf_real}
  \check{\bx}_{\mathrm{R,ZF}}=\bH_{\mathrm{R,S}}^{\dagger}\hat{\by}_{\mathrm{R}}.
\end{equation}
With $\bH$ and $\hat{\by}$ which are defined in Section \ref{sec2}, the same receiver with \eqref{zf_real} can be implemented in the complex domain as
\begin{equation}\label{zf_complex}
  \check{\bx}_{\mathrm{ZF}}=\bH^{\dagger}\hat{\by}
\end{equation}
where $\check{\bx}_{\mathrm{ZF}}\in \mathbb{C}^{N_t}$.  The elements of $\check{\bx}_{\mathrm{ZF}}$ may not be in the $M$-ary constellation $\cS$ used to generated the transmitted vector $\bx$.  Thus, the fusion center needs to estimate $\hat{\bx}_{\mathrm{ZF}}=\begin{bmatrix}\hat{x}_{\mathrm{ZF},1} & \hat{x}_{\mathrm{ZF},2} & \cdots & \hat{x}_{\mathrm{ZF},N_t}\end{bmatrix}^T$ by selecting the closest constellation point from $\check{\bx}_{\mathrm{ZF}}=\begin{bmatrix}\check{x}_{\mathrm{ZF},1} & \check{x}_{\mathrm{ZF},2} & \cdots & \check{x}_{\mathrm{ZF},N_t}\end{bmatrix}^T$ as
\begin{equation}\label{zf_est}
  \hat{x}_{\mathrm{ZF},n}=\argmin_{s'\in \cS}|\check{x}_{\mathrm{ZF},n}-s'|^2
\end{equation}
for $n=1,\cdots,N_t$.  The complexity of the ZF-type receiver is much lower than that of the ML receiver because each of these minimizations is over a set of $M$ elements.

\subsection{Receiver performance}\label{analysis}
In this section, we analyze the performance of ML and ZF-type estimators, which are suboptimal than the proposed receivers.  The following lemma shows the behavior of the ML estimator in the asymptotic regime of $K$ for arbitrary $\rho>0$.

\begin{lemma}\label{CONVERGE_ML}
Let $\check{\bx}_{\mathrm{ML}}$ be the outcome of the ML estimator
\begin{equation}\label{ml_hatx}
\check{\bx}_{\mathrm{ML}} = \argmax_{\substack{\bx'\in \mathbb{C}^{N_t},\\ \|\bx'\|^2=N_t}}L(\bx').
\end{equation}
For arbitrary $\rho>0$, $\check{\bx}_{\mathrm{ML}}$ converges to the true transmitted vector $\bx$ in probability, i.e.,
\begin{equation*}
\check{\bx}_{\mathrm{ML}} \stackrel{p}{\longrightarrow} \bx
\end{equation*}
as $K\rightarrow \infty$.
\end{lemma}
\begin{IEEEproof}
We consider the real domain in the proof to simplify notation.  The lemma can be proved by showing the inequality
\begin{equation*}
L(\bx_{\mathrm{R}})> L(\bu_{\mathrm{R}})
\end{equation*}
in probability for any $\bu_{\mathrm{R}}\in \mathbb{R}^{2N_t}\setminus \{\bx_{\mathrm{R}}\}$ with the constraint $\|\bu_{\mathrm{R}}\|^2=N_t$ when $K\rightarrow \infty$ for arbitrary $\rho>0$.  We take logarithm of the likelihood function and have
\begin{align*}
  \log L(\bx^{\ddag}) & = \log \left(\prod_{i=1}^{2}\prod_{k=1}^{K}\Phi\left(\sqrt{\frac{2\rho}{N_t}}\widetilde{\bh}_{\mathrm{R},k,i}^T\bx^{\ddag}\right)\right)\\
  & = \sum_{i=1}^{2}\sum_{k=1}^{K}\log \Phi\left(\sqrt{\frac{2\rho}{N_t}}\widetilde{\bh}_{\mathrm{R},k,i}^T\bx^{\ddag}\right).
\end{align*}
Because the $\widetilde{\bh}_{\mathrm{R},k,i}$'s are independent for all $k$,
\begin{align*}
  &\lim_{K\rightarrow \infty}\frac{1}{K}\sum_{k=1}^{K}\log \Phi\left(\sqrt{\frac{2\rho}{N_t}}\widetilde{\bh}_{\mathrm{R},k,i}^T\bx^{\ddag}\right)\\
  &\qquad \qquad \qquad \stackrel{p}{\longrightarrow} \mathrm{E}\left[\log \Phi\left(\sqrt{\frac{2\rho}{N_t}}\widetilde{\bh}_{\mathrm{R},k,i}^T\bx^{\ddag}\right)\right]
\end{align*}
by the weak law of large numbers, and we have
\begin{align*}
  \frac{1}{K}\log L(\bx^{\ddag}) & \stackrel{p}{\longrightarrow} 2\mathrm{E}\left[\log \Phi\left(\sqrt{\frac{2\rho}{N_t}}\widetilde{\bh}_{\mathrm{R},k,i}^T\bx^{\ddag}\right)\right]
\end{align*}
as $K\rightarrow \infty$ where the expectation is taken over the channel.

Then, we need to show that
\begin{equation*}
  \mathrm{E}\left[\log \Phi\left(\sqrt{\frac{\rho}{N_t}}\widetilde{\bh}_{\mathrm{R},k,i}^T\bx_{\mathrm{R}}\right)\right]> \mathrm{E}\left[\log \Phi\left(\sqrt{\frac{\rho}{N_t}}\widetilde{\bh}_{\mathrm{R},k,i}^T\bu_{\mathrm{R}}\right)\right]
\end{equation*}
where the expectations are taken over the channel.  Because $\log \Phi(\cdot)$ is a strictly monotonically increasing concave function, the above inequality is true if $\widetilde{\bh}_{\mathrm{R},k,i}^T\bx_{\mathrm{R}}$ first-order stochastically dominates $\widetilde{\bh}_{\mathrm{R},k,i}^T\bu_{\mathrm{R}}$ \cite{stochastic_dominance}.  In Appendix \ref{sto_domi_proof}, we show
\begin{equation}\label{sto_domin}
  \widetilde{\bh}_{\mathrm{R},k,i}^T\bx_{\mathrm{R}}\stackrel{d}{>}\widetilde{\bh}_{\mathrm{R},k,i}^T\bu_{\mathrm{R}}
\end{equation}
conditioned on the received signal $y_{\mathrm{R},k,i}$ where $\stackrel{d}{>}$ denotes strict first-order stochastic dominance.
\end{IEEEproof}

We define the MSE between $\bx$ and $\check{\bx}$ as
\begin{equation*}
  \mathrm{MSE} = \frac{1}{N_t}\mathrm{E}\left[\|\bx-\check{\bx}\|^2\right]
\end{equation*}
where the expectation is taken over the realizations of channel and noise.  The following corollary shows the MSE performance of the ML estimator in the asymptotic regime of $K$ for arbitrary $\rho>0$.
\begin{corollary}
The MSE of the ML estimator converges to zero, i.e.,
\begin{equation*}
\lim_{K \rightarrow \infty}\mathrm{MSE_{ML}}\rightarrow 0
\end{equation*}
for arbitrary $\rho>0$.
\end{corollary}
\begin{IEEEproof}
Note that the norm of $\bx$ is bounded, i.e.,
\begin{equation*}
  \|\bx\|^2=N_t<\infty.
\end{equation*}
The convergence in probability of a random variable with a bounded norm implies the convergence in mean-square sense \cite{Papou}.  Thus, we have
\begin{equation*}
  \lim_{K\rightarrow \infty} \mathrm{E}\left[\|\bx-\check{\bx}_{\mathrm{ML}}\|^2\right] \rightarrow 0,
\end{equation*}
which finishes the proof.
\end{IEEEproof}

This analytical derivation shows that the proposed ML receiver (which should perform better than the ML estimator) can decode the transmitted vector without any error with large $K$ and fixed $\rho$.  Moreover, numerical studies in Section \ref{simulation} show that increasing $\rho$ would be sufficient for the ML receiver to decode the transmitted vector correctly with fixed, but sufficiently large, $K$.

We now analyze the MSE of the ZF-type estimator.  Although it is difficult to derive the MSE of the ZF-type estimator in general, we are able to have a closed-form expression for $\mathrm{MSE}_{\mathrm{ZF}}$ by approximating quantization loss as additional Gaussian noise where the approximation is frequently adopted in many frame expansion works, e.g., \cite{frame_exp1,frame_exp2,frame_exp4}.
\begin{lemma}\label{zf_anal}
If we approximate the quantization error using an additional Gaussian noise $\bw$ as
\begin{equation}\label{quant_approx}
  \hat{\by}= \sqrt{\frac{\rho}{N_t}}\bH\bx + \bn+\bw,
\end{equation}
with\footnote{We normalize the covariance of $\bw$ with $\frac{\rho}{N_t}$ to have statistically equal quantization loss regardless of the received SNR.} $\bw\sim \cC\cN(\mathbf{0}_K,\sigma^2_q \frac{\rho}{N_t}\bI_K)$ and assume $\bH^H\bH = K\mathbf{I}_{N_t}$, the MSE of the ZF-type estimator is given as
\begin{equation*}
  \mathrm{MSE}_{\mathrm{ZF}} = \mathrm{E}\left[\|\bx-\check{\bx}_{\mathrm{ZF}}\|^2\right]=\frac{N_t\rho^{-1}+\sigma_q^2}{K}
\end{equation*}
where $\check{\bx}_{\mathrm{ZF}}$ is defined in \eqref{zf_complex}.
\end{lemma}
\begin{IEEEproof}
Because $\bn$ and $\bw$ are independent, \eqref{quant_approx} can be rewritten as
\begin{equation*}
  \hat{\by}= \sqrt{\frac{\rho}{N_t}}\bH\bx + \bn'
\end{equation*}
with $\bn'\sim \cC\cN\left(\mathbf{0}_K,\left(1+\sigma^2_q \frac{\rho}{N_t}\right) \bI_K \right)$.  It was shown by Proposition 1 in \cite{frame_exp1} that $\mathrm{MSE}_{\mathrm{ZF}}$ can be bounded as
\begin{equation}\label{lower_upper_mse}
  \frac{K\left(N_t\rho^{-1}+\sigma^2_q\right)}{B^2}\leq \mathrm{MSE}_{\mathrm{ZF}}\leq \frac{K\left(N_t\rho^{-1}+\sigma^2_q\right)}{A^2}
\end{equation}
where $A$ and $B$ are fixed constants that satisfy
\begin{equation*}
  A\mathbf{I}_{N_t}\leq \bH^H \bH \leq B\mathbf{I}_{N_t}.
\end{equation*}
The matrix inequality $A\mathbf{I}_{N_t}\leq \bH^H \bH$ means that the matrix $\bH^H \bH-A\mathbf{I}_{N_t}$ is a positive semidefinite matrix.  Due to the assumption on the channel matrix, we have
\begin{equation*}
  A=B=K,
\end{equation*}
and the lower and upper bounds in \eqref{lower_upper_mse} both become $\frac{\left(N_t\rho^{-1}+\sigma^2_q\right)}{K}$, which finishes the proof.
\end{IEEEproof}

Note that the assumption $\bH^H\bH = K\mathbf{I}_{N_t}$ in Lemma \ref{zf_anal} can be satisfied with a large number of receive nodes because
\begin{equation*}
  \bH^H \bH \stackrel{p}{\longrightarrow} K\mathbf{I}_{N_t}
\end{equation*}
as $K\rightarrow \infty$ under our $\cC\cN(0,1)$ i.i.d. channel assumption.  Moreover, we have the following corollary when $\rho$ becomes large.

\begin{corollary}\label{zf_anal_large_rho}
With the same assumptions used in Lemma \ref{zf_anal}, the MSE of the ZF-type estimator is given as
\begin{equation*}
  \mathrm{MSE}_{\mathrm{ZF}} = \frac{\sigma_q^2}{K}
\end{equation*}
when $\rho$ goes to infinity.
\end{corollary}
\begin{IEEEproof}
The proof of Corollary \ref{zf_anal_large_rho} is a direct consequence of taking the limit $\rho \rightarrow \infty$ on the  result of Lemma \ref{zf_anal}.
\end{IEEEproof}

Lemma \ref{zf_anal} and Corollary \ref{zf_anal_large_rho} show that we can make $\mathrm{MSE_{ZF}}$ arbitrarily small by increasing $K$ regardless of the effect of noise or quantization error.  However, due to the quantization process at each receive node, we have $\sigma_q^2>0$, and $\mathrm{MSE}_{\mathrm{ZF}}$ never goes to zero with fixed $K$ even when $\rho \rightarrow \infty$, which gives an error rate floor in the high SNR regime.  These MSE analyses are based on the ZF-type estimator and the approximation of the quantization process in \eqref{quant_approx}; however, the numerical results in Section \ref{simulation} show that the analyses also hold for the SER case with actual quantization process using the proposed ZF-type receiver.
\begin{figure*}
\centering
\subfloat[$\rho=10$.]{
\includegraphics[width=1\columnwidth]{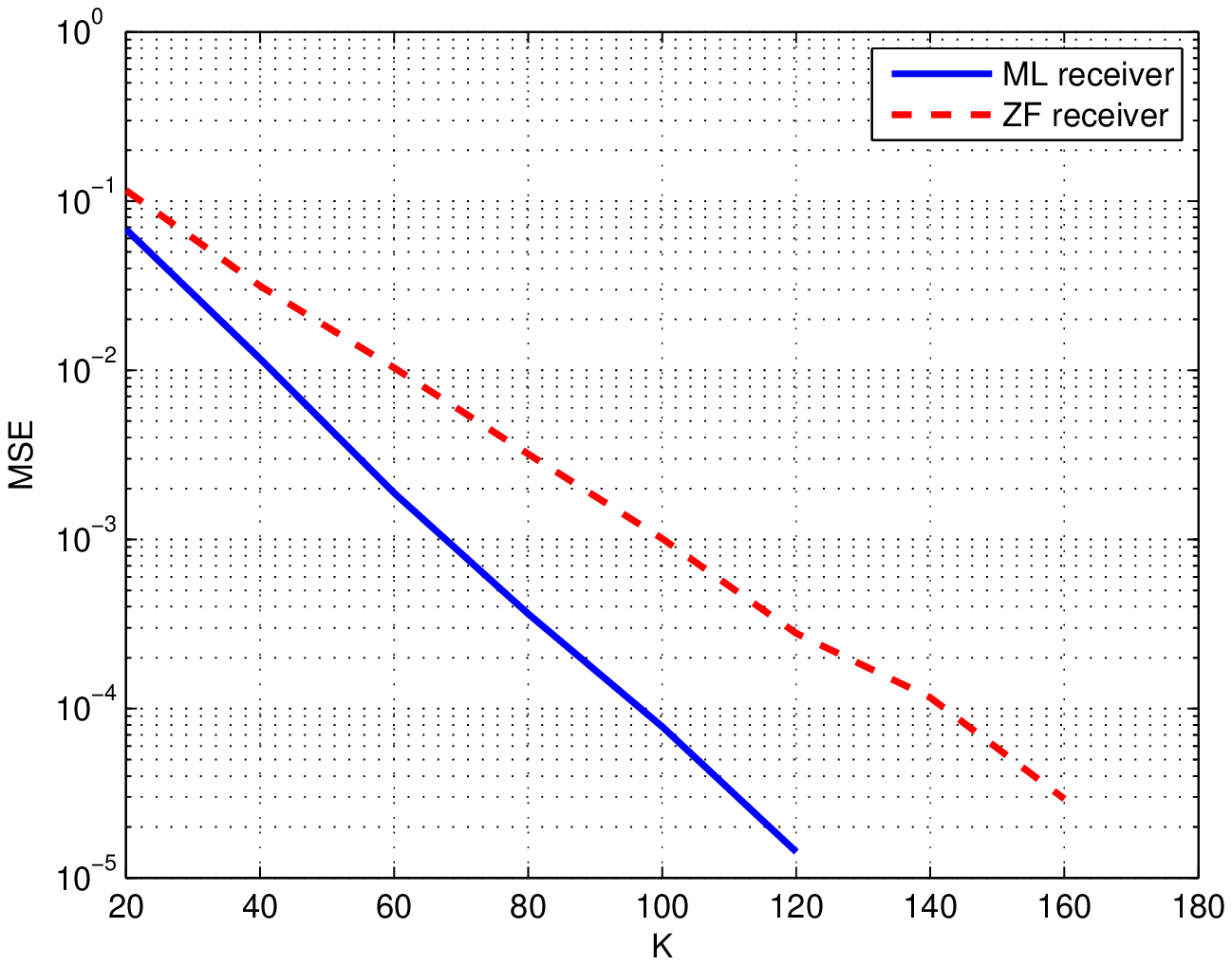}
\label{mse1}
}
\subfloat[$K=50$.]{
\includegraphics[width=1\columnwidth]{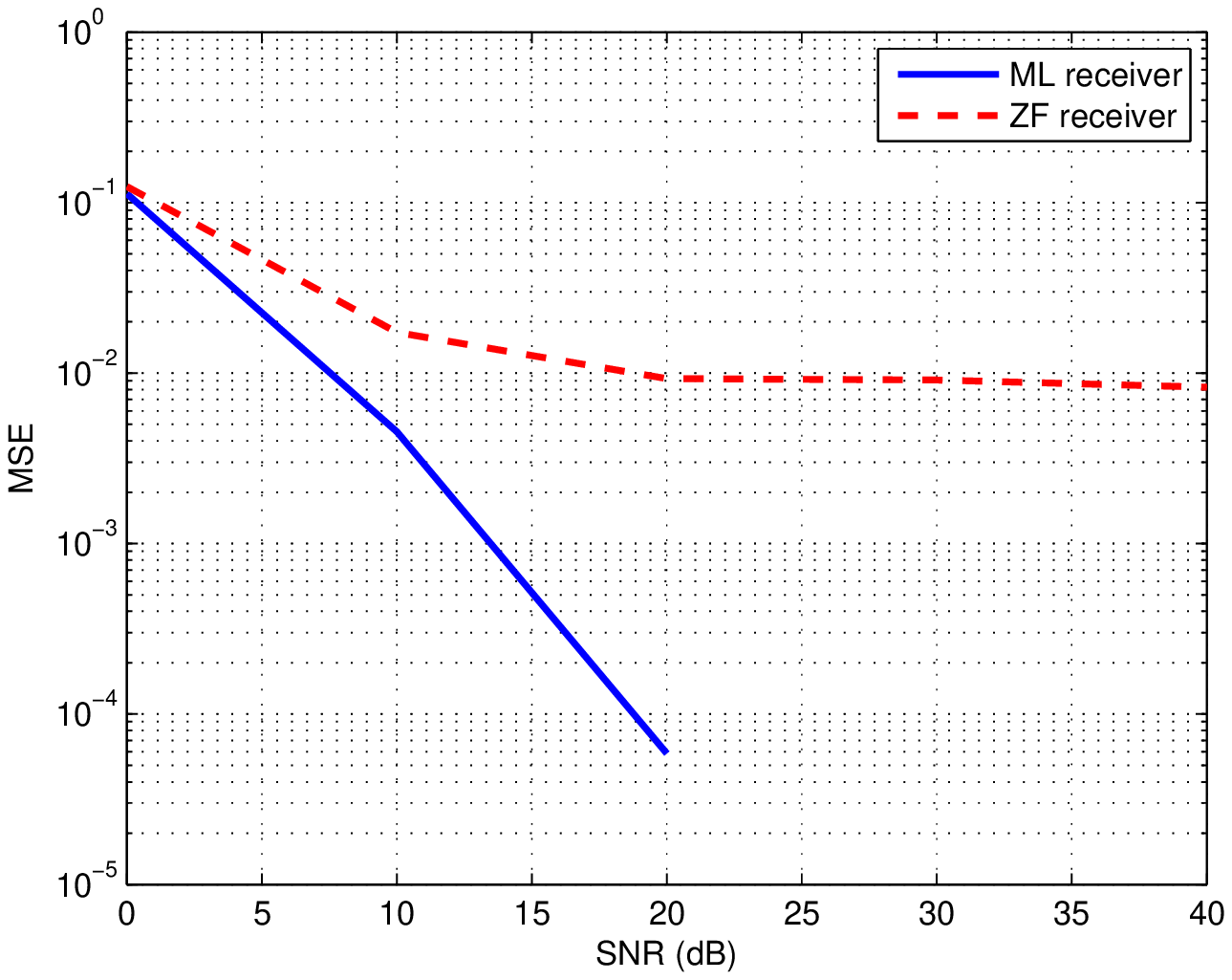}
\label{mse2}
}
\caption{The MSE of the ML and the ZF-type receiver with increasing either of $K$ or $\rho$.  We set $M=8$ (8PSK) and $N_t=4$ for both figures.}
\label{mse_fig}
\end{figure*}

\subsection{Modified zero-forcing-type receiver}\label{zf_mod}
As mentioned in the previous subsection, the ZF-type receiver suffers from an error rate floor when $\rho$ goes to infinity with fixed $K$.  Although the error rate floor is indeed inevitable with the ZF-type receiver, we can improve the SER of the ZF-type receiver in the high SNR regime by performing post-processing for $\hat{\bx}_{\mathrm{ZF}}$ given in \eqref{zf_est}.

When $\rho\rightarrow \infty$, the effect of noise disappears, and we have
\begin{equation*}
  \widetilde{\bH}_{\mathrm{R,S}}\bx_{\mathrm{R}}\succeq \mathbf{0}_{2K}
\end{equation*}
by the sign adjustment, where $\widetilde{\bH}_{\mathrm{R,S}}$ is defined in \eqref{H_tilde_def}, $\bx_{\mathrm{R}}$ is the transmitted vector in the real domain, and $\succeq$ represents element-wise inequality.  Even in the high SNR regime, however, the $\hat{\bx}_{\mathrm{ZF}}$ that is estimated from the ZF-type receiver may not satisfy the inequality constraints, which would cause an error rate floor.  Thus, we formulate a linear program as
\begin{align*}
  &\max_{\hat{\bx}_{\mathrm{R}}\in \mathbb{R}^{2N_t}} \hat{\bx}_{\mathrm{ZF}}^T \hat{\bx}_{\mathrm{R}} \\
  &\mathrm{s.t.} \quad \widetilde{\bH}_{\mathrm{R,S}}\hat{\bx}_{\mathrm{R}}\succeq \mathbf{0}_{2K}
\end{align*}
to force the estimate $\hat{\bx}_{\mathrm{R}}$ to satisfy the inequality constraints.  The estimate $\hat{\bx}_{\mathrm{R}}$ should be mapped to $\cS$ as in \eqref{zf_est} before decoding.

It was shown in \cite{frame_exp1} that in the context of frame expansion without any noise, the reconstruction method by linear programming can give a MSE proportional to $\frac{1}{K^2}$, which is much better than the ZF-type receiver which results in a MSE proportional to $\frac{1}{K}$.  However, if $\rho$ is not large enough, this post-processing by linear programming can cause performance degradation because the sign refinement may not be perfect, resulting in incorrect inequality constraints for the linear programming.  Moreover, in this case, having more receive nodes may cause more errors due to the higher chance of having wrong inequality constraints.  Note that more receive nodes corresponds to more rows in $\widetilde{\bH}_{\mathrm{R,S}}$ that force more inequality constraints.  We numerically evaluate the modified ZF-type receiver in Section \ref{simulation}.


\section{Numerical Results}\label{simulation}

\begin{figure*}
\centering
\subfloat[$M=4$ (QPSK) constellation for $\cS$ and $N_t=6$.]{
\includegraphics[width=1\columnwidth]{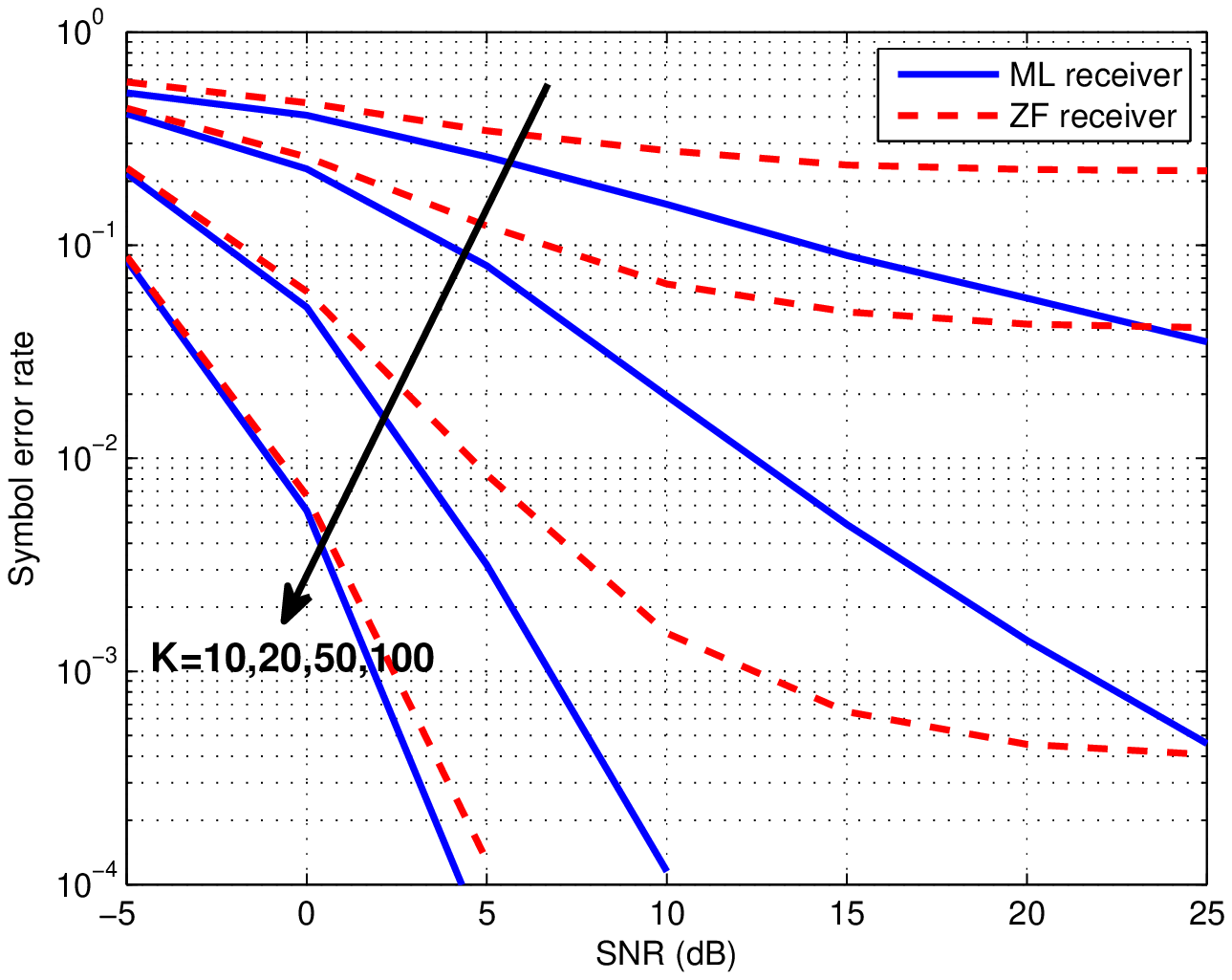}
\label{M4Nt6}
}
\subfloat[$M=8$ (8PSK) constellation for $\cS$ and $N_t=4$.]{
\includegraphics[width=1\columnwidth]{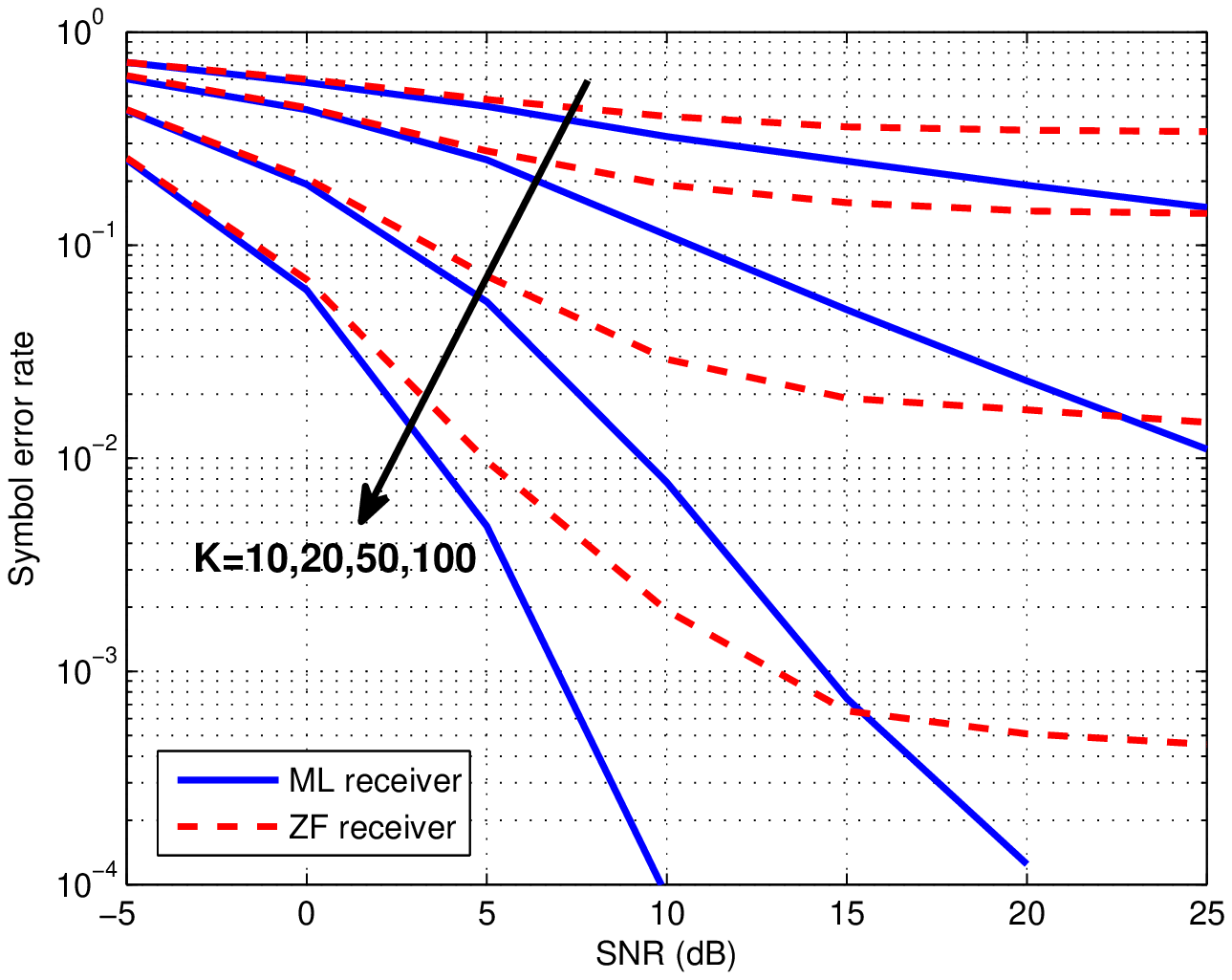}
\label{M8Nt4}
}
\caption{Symbol error rate (SER) vs. SNR in dB scale with different values of $N_t$ and $M$ for the constellation $\cS$. Both figures are the case of 12 bits transmission per channel use.}
\label{ser_fig}
\end{figure*}

In this section, we evaluate the proposed receivers with Monte Carlo simulations.  We first evaluate the MSE of the ML and the ZF-type receiver with 10000 channel realizations to verify the analytical results derived in Section \ref{analysis}.  In Fig. \ref{mse1}, we increase $K$ with fixed $\rho=10$ (i.e., an SNR of 10 dB\footnote{Recall that $\rho$ is related to total transmit power, not per antenna transmit power, in our system setup.}).  In Fig. \ref{mse2}, we fix $K=50$ and increase $\rho$.  We set $N_t=4$ and $M=8$ (8PSK for $\cS$) in both figures.  It is clear that the MSE of the ML receiver decreases without bound if either of $K$ or $\rho$ becomes larger while the MSE of the ZF-type receiver is certainly bounded with fixed $K$ as $\rho$ becomes larger.  However, if $K$ becomes larger, the MSE of the ZF-type receiver also decreases without bound.

To see the diversity gain of each receiver, we consider the average SER which is defined as
\begin{equation*}
  \mathrm{SER} = \frac{1}{N_t}\sum_{n=1}^{N_t}\mathrm{E}\left[\mathrm{Pr}\left(\hat{x}_n\neq x_n\mid \bx~\text{sent},\bH,\bn,\rho,N_t,K,\cS \right)\right]
\end{equation*}
where the expectation is taken over $\bx$, $\bH$, and $\bn$.  We compare the SERs of ML and ZF-type (without the modification by the linear programming) receivers regarding the transmit SNR $\rho$ in dB scale with different values of $N_t$ and $M$ in Fig. \ref{ser_fig}.  Note that both figures are for the case of 12 bits transmission per channel use because the total number of bits transmitted per channel use is given as
\begin{equation*}
   B_{\mathrm{tot}}=N_t\log_2 M.
\end{equation*}
It is clear from the figures that as $\rho$ or $K$ increase, the SER of the ML receiver becomes smaller without any bound while that of the ZF-type receiver is certainly bounded in the high SNR regime.  However, the SER of the ZF-type receiver can be improved by increasing $K$, which is the same as the MSE results.  The results show that the ZF-type receiver would be a good option for distributed reception with a large number of receive nodes in the IoT environment.
\begin{figure}[t]
  \centering
  \includegraphics[width=1\columnwidth]{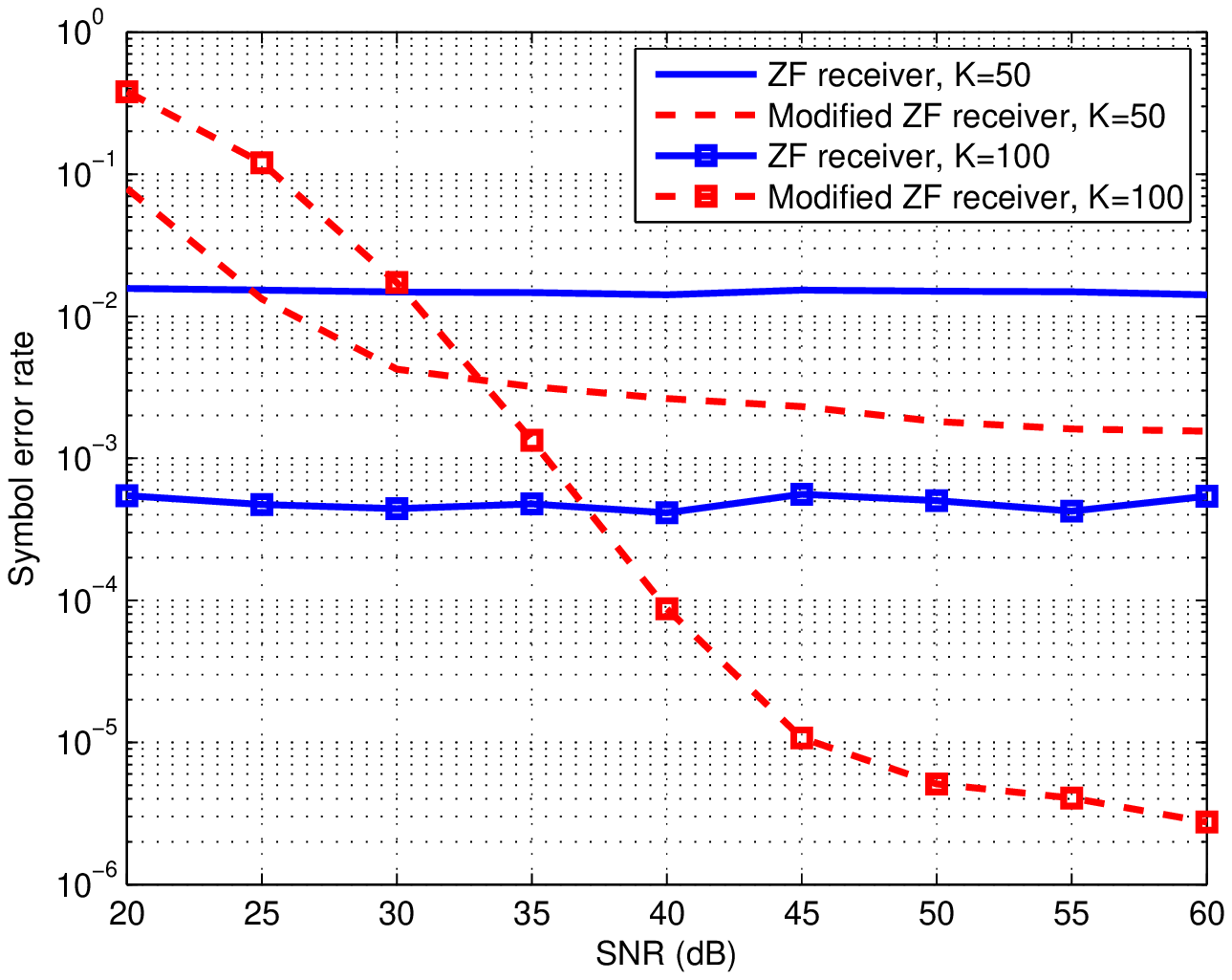}\\
  \caption{Symbol error rate (SER) vs. SNR in dB scale for the ZF-type and modified ZF-type receivers with $N_t=4$ and $M=8$ for the constellation $\cS$.}\label{mzf_eval}
\end{figure}
\begin{figure}[t]
  \centering
  \includegraphics[width=1\columnwidth]{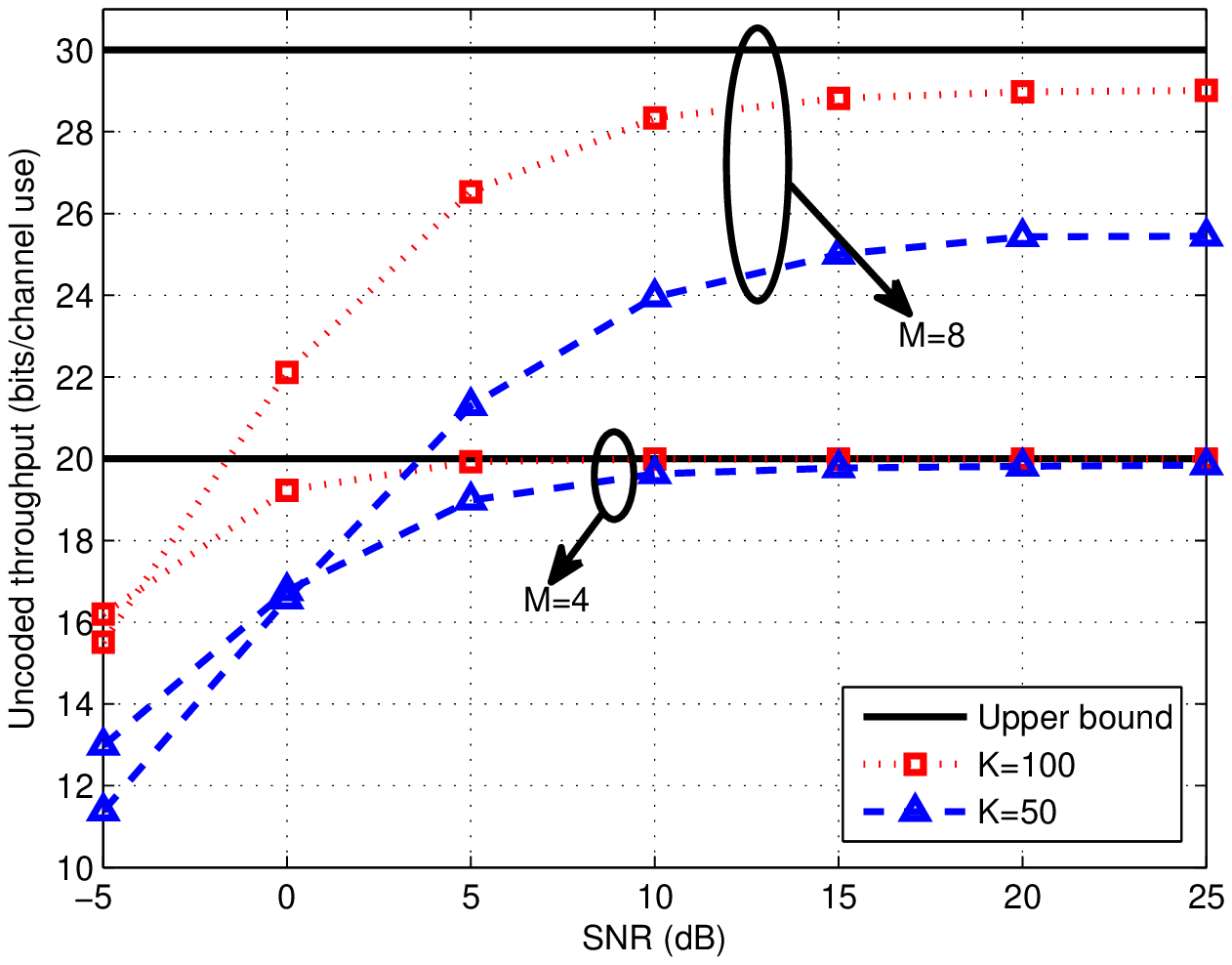}\\
  \caption{Uncoded throughput of the ZF-type receiver vs. SNR in dB scale with $N_t=10$ and different values of $K$ and $M$.}\label{goodput}
\end{figure}

Comparing these figures, if the number of transmit antennas $N_t$ at the transmitter is large, it is desirable to simultaneously transmit more symbols chosen from a smaller sized constellation to get better SER results when the number of transmitted bits per channel use, $B_{\mathrm{tot}}$, is fixed for both the ML and ZF-type receivers.  This result is suitable to massive MIMO systems where the transmitter is equipped with a large number of antennas.

In Fig. \ref{mzf_eval}, we plot the SERs of the ZF-type receiver and a ZF-type receiver modified to use linear programming explained in Section \ref{zf_mod}.  We only consider the high SNR regime because the modified ZF-type receiver is aimed to increase the performance of the ZF-type receiver when the SNR is high. The figure clearly shows that the modified ZF-type receiver performs much better than the ZF-type receiver when the effect of noise becomes negligible; however, it performs worse than the ZF-type receiver when the SNR is not sufficiently high.  Moreover, having more receive nodes deteriorates the performance of the modified ZF-type receiver in this case, which is explained in Section \ref{zf_mod}.

To evaluate the benefit of spatial multiplexing in distributed reception, we plot the uncoded throughput versus SNR of the ZF-type\footnote{We do not simulate the ML receiver in this scenario due to the excessive complexity of the ML receiver when $N_t=10$.} receiver without the post-processing by the linear programming with $N_t=10$ in Fig. \ref{goodput}, where uncoded throughput is the number of successfully transmitted bits per channel use and defined as
\begin{equation*}
  B_{\mathrm{tot}}(1-\mathrm{SER}).
\end{equation*}
For $M=4$, i.e., when the transmitter transmits QPSK symbols, uncoded throughput of $K=50$ and $K=100$ cases both approach their upper bound of 20 bits transmission per channel use as $\rho$ becomes large.  On the other hand, when $M=8$ and the transmitter adopts 8PSK symbols, there is a certain gap between the upper bound and the case with a finite value of $K$.  Note that the gap will become smaller as $K$ gets larger.

It is interesting to point out that there is a crossover point between the plots of $M=4$ and $M=8$ with the same $K$.  When the SNR is low, it would be better for the transmitter to use a small size constellation to provide a lower SER and higher uncoded throughput.  However, if the SNR is larger, then a good strategy for the transmitter would be to use high order constellation for high data rate transmission.

\section{Conclusion}\label{conclusion}
In this paper, we studied a distributed reception scenario where the transmitter is equipped with multiple transmit antennas and broadcasts multiple independent data symbols by spatial multiplexing to a set of geographically separated receive nodes through fading channels.  Each receive node then processes its received signal and forwards it to the fusion center, and the fusion center tries to decode the transmitted data symbols by exploiting the forwarded information and global channel knowledge.  We implemented an optimal ML receiver and a low-complexity ZF-type receiver for this scenario.  The SER of the ML receiver can be made arbitrarily small by increasing SNR and the number of receive nodes.  The ZF-type receiver suffers from an error rate floor as the SNR increases.  This floor can be lowered by increasing the number of receive nodes.

The scenario studied in this paper, i.e., high data rate transmission by spatial multiplexing in distributed reception, may become popular in the near future with the emergence of the Internet of Things (IoT) where we can easily have a numerous number of receive nodes.  To make the scenario more practical, the fusion center may decode the transmitted symbols with partial or no global channel knowledge.  Extending our framework to non-PSK constellations is an interesting future research topic.

\appendices
\section{}\label{phi_lemma}
\begin{lemma*}
For arbitrary $s$ and $c$ that satisfy $s>c>0$, we have
\begin{equation*}
  \left(\Phi(s)\right)^2>\Phi(s+c)\Phi(s-c).
\end{equation*}
\end{lemma*}
\begin{IEEEproof}
With $s>c>0$, we have the inequality
\begin{equation*}
\Phi(s)-\Phi(s-c)>\Phi(s+c)-\Phi(s).
\end{equation*}
Then, we have
\begin{equation*}
2\Phi(s)>\Phi(s+c)+\Phi(s-c)
\end{equation*}
which is equivalent to
\begin{align*}
4\left(\Phi(s)\right)^2&>\left(\Phi(s+c)+\Phi(s-c)\right)^2\\
&=\left(\Phi(s+c)\right)^2+\left(\Phi(s-c)\right)^2+2\Phi(s+c)\Phi(s-c)\\
&\stackrel{(a)}{\geq} 4\Phi(s+c)\Phi(s-c)
\end{align*}
where $(a)$ is because
\begin{equation*}
\left(\Phi(s+c)-\Phi(s-c)\right)^2\geq 0,
\end{equation*}
which finishes the proof.
\end{IEEEproof}

\section{Proof of First-Order Stochastic Dominance}\label{sto_domi_proof}
We drop unnecessary subscripts to simplify notation.  Recall that
\begin{align*}
  y&=\sqrt{\frac{\rho}{N_t}}\bh^T\bx+n,\\
  \widetilde{\bh}&=\hat{y}\bh
\end{align*}
where $\hat{y}=\mathrm{sgn}(y)$.  Using the fact that $\bh$ is rotationally invariant, we assume the transmitted vector is given as\footnote{Because we consider the ML estimator not receiver in this proof, we do not have to restrict the elements of $\bx$ from PSK constellation points.} $\bx=\begin{bmatrix}\sqrt{N_t} & 0 &\cdots &0\end{bmatrix}^T$.  Then, we have
\begin{equation*}
  y=\sqrt{\rho}h_1+n.
\end{equation*}
Because $y\sim \cN(0,\frac{\rho+1}{2})$ and $n\sim\cN(0,\frac{1}{2})$, the distribution of $\sqrt{\rho}h_1$ conditioned on $y$ is $\cN(\mu,\gamma^2)$ where
\begin{align*}
  \mu=\frac{\rho}{\rho+1}y,\quad \gamma^2 & = \frac{\rho}{2(\rho+1)}.
\end{align*}
Let $c=\frac{\rho}{\rho+1}$.  Then, we can write $\sqrt{\rho}h_1=cy+w$ where $w\sim \cN(0,\gamma^2)$.  Moreover, we have
\begin{align*}
  \sqrt{\frac{\rho}{N_t}}\widetilde{\bh}^T\bx&=\sqrt{\rho}\hat{y}h_1=c|y|+\hat{y}w\stackrel{d}{=}|cy|+w
\end{align*}
conditioned on $y$ where the third equality comes from the independence of $\hat{y}$ and $w$.  Note that $\stackrel{d}{=}$ denotes stochastic equivalence.

Now we want to compute the distribution of $\sqrt{\frac{\rho}{N_t}}\widetilde{\bh}^T\bu$ for a fixed $\bu$ given $y$.  Note that
\begin{equation*}
  \bh^T\bu=\sum_{i=1}^{2N_t}h_iu_i\stackrel{d}{=}u_1h_1+z\sqrt{N_t-u_1^2}
\end{equation*}
where $z\sim \cN(0,\frac{1}{2})$.  Then, we have
\begin{align*}
  \sqrt{\frac{\rho}{N_t}}\widetilde{\bh}^T\bu&\stackrel{d}{=}\frac{u_1}{\sqrt{N_t}}(c|y|+w)+\hat{y}z\sqrt{\rho\left(1-\frac{u_1^2}{N_t}\right)}\\
  &\stackrel{d}{=}\frac{u_1}{\sqrt{N_t}}(c|y|+w)+z\sqrt{\rho\left(1-\frac{u_1^2}{N_t}\right)}\\
  & = u(c|y|+w)+z\sqrt{\rho\left(1-u^2\right)}
\end{align*}
where the second equality is due to the independence of $\hat{y}$ and $z$ and the third equality comes from the variable substitution $u=\frac{u_1}{\sqrt{N_t}}$.  Note that $-1\leq u<1$.  If $u=1$, then $\bu$ becomes $\bx$, which violates our assumption.

We now break up $uw+z\sqrt{\rho\left(1-u^2\right)}$ into two independent zero-mean Gaussian random variables $v_1$ and $v_2$ where
\begin{align*}
  v_1 & \sim \cN\left(0,(1-u^2)\frac{\rho^2}{2(\rho+1)}\right),\quad v_2 \sim \cN(0,\gamma^2).
\end{align*}
Finally, for a given $y$, we have
\begin{align}
\nonumber  \sqrt{\frac{\rho}{N_t}}\widetilde{\bh}^T\bu &\stackrel{d}{=}u(c|y|+w)+z\sqrt{\rho\left(1-u^2\right)} \\
\nonumber  & = uc|y|+v_1+v_2 \\
& \stackrel{d}{<} |uc|y|+v_1|+v_2\label{strict_dominance}\\
\nonumber  & = |ucy+\hat{y}v_1|+v_2 \\
  & \stackrel{d}{=} |ucy+v_1|+v_2\label{dist_equal1} \\
  & \stackrel{d}{=}|cy|+w\label{dist_equal2}\\
\nonumber  & \stackrel{d}{=} \sqrt{\frac{\rho}{N_t}}\widetilde{\bh}^T\bx.
\end{align}
To show the strict stochastic dominance in \eqref{strict_dominance}, recall that $uc|y|$ is a fixed number given $y$, and $v_1$ is a Gaussian random variable.  Thus, the complementary cumulative distribution function of $|uc|y|+v_1|$ should be strictly greater than that of $uc|y|+v_1$.  The stochastic equivalence in \eqref{dist_equal1} is because $\hat{y}$ and $v_1$ are independent and \eqref{dist_equal2} is due to the facts that
\begin{align*}
  |u|cy+v_1 & \sim \cN\left(0,\frac{\rho^2}{2(\rho+1)}\right) \\
  cy & \sim \cN\left(0,\frac{\rho^2}{2(\rho+1)}\right)
\end{align*}
and $v_2\stackrel{d}{=} w$.  Thus, \eqref{sto_domin} holds, and we have the claim.

\bibliographystyle{IEEEtran}
\bibliography{refs}

\end{document}